\documentclass[12pt]{iopart}
\usepackage[english]{babel}
\usepackage[T1]{fontenc}
\usepackage[colorlinks=true,linkcolor=blue,citecolor=blue,urlcolor=blue]{hyperref}
\expandafter\let\csname equation*\endcsname\relax
\expandafter\let\csname endequation*\endcsname\relax
\usepackage{bm,bbm,amssymb,amsmath}
\usepackage{graphicx}
\usepackage{multirow,booktabs,dcolumn}
\usepackage{multirow,dcolumn}
\usepackage{rotating}
\usepackage{slashed}
\usepackage{isotope}
\usepackage{ifthen}
\usepackage[capitalize]{cleveref}
\usepackage{nicefrac}
\usepackage{ulem}

\usepackage[sort&compress,numbers,square]{natbib}

\newcommand{\eu}[1]{\mathrm{e}^{#1}}
\newcommand{\ii}{\mathrm{i}}

\makeatletter
\DeclareRobustCommand{\cev}[1]{%
  \mathpalette\do@cev{#1}%
}
\newcommand{\do@cev}[2]{%
  \fix@cev{#1}{+}%
  \reflectbox{$\m@th#1\vec{\reflectbox{$\fix@cev{#1}{-}\m@th#1#2\fix@cev{#1}{+}$}}$}%
  \fix@cev{#1}{-}%
}
\newcommand{\fix@cev}[2]{%
  \ifx#1\displaystyle
    \mkern#23mu
  \else
    \ifx#1\textstyle
      \mkern#23mu
    \else
      \ifx#1\scriptstyle
        \mkern#22mu
      \else
        \mkern#22mu
      \fi
    \fi
  \fi
}

\makeatother

\newcommand{\normord}[1]{:\mathrel{\mkern2mu #1 \mkern2mu}:}

\newcommand{\nxlo}[1]{%
   \ifnum0=#1\relax%
      LO%
   \else%
      \ifnum1=#1\relax%
         NLO%
      \else%
         N$^{#1}$LO%
      \fi
   \fi
}

\begin{document}

\title[\textit{Ab initio} short-range-correlation scaling factors]{\textit{Ab initio} short-range-correlation scaling factors from light to medium-mass nuclei}

\author{J. E. Lynn}
\address{Institut f\"ur Kernphysik, Technische Universit\"at Darmstadt, 64289 Darmstadt, Germany}
\address{ExtreMe Matter Institute EMMI, GSI Helmholtzzentrum f\"ur Schwerionenforschung GmbH, 64291 Darmstadt, Germany}

\author{D. Lonardoni}
\address{Facility for Rare Isotope Beams, Michigan State University, East Lansing, Michigan 48824, USA}
\address{Theoretical Division, Los Alamos National Laboratory, Los Alamos, New Mexico 87545, USA}
\ead{lonardoni@nscl.msu.edu}

\author{J. Carlson}
\address{Theoretical Division, Los Alamos National Laboratory, Los Alamos, New Mexico 87545, USA}

\author{J.-W. Chen}
\address{Department of Physics, CTP and LeCosPA, National Taiwan University, Taipei 10617, Taiwan}
\address{Center for Theoretical Physics, Massachusetts Institute of Technology, Cambridge, MA 02139, USA}

\author{W. Detmold}
\address{Center for Theoretical Physics, Massachusetts Institute of Technology, Cambridge, MA 02139, USA}

\author{S. Gandolfi}
\address{Theoretical Division, Los Alamos National Laboratory, Los Alamos, New Mexico 87545, USA}

\author{A. Schwenk}
\address{Institut f\"ur Kernphysik, Technische Universit\"at Darmstadt, 64289 Darmstadt, Germany}
\address{ExtreMe Matter Institute EMMI, GSI Helmholtzzentrum f\"ur Schwerionenforschung GmbH, 64291 Darmstadt, Germany}
\address{Max-Planck-Institut f\"ur Kernphysik, Saupfercheckweg 1, 69117 Heidelberg, Germany}

\begin{abstract}
High-energy scattering processes, such as deep inelastic scattering
(DIS) and quasielastic (QE) scattering provide a wealth of information
about the structure of atomic nuclei.
The remarkable discovery of the empirical linear relationship between
the slope of the European Muon Collaboration (EMC) effect in DIS and the short-range-correlation
(SRC) scaling factors $a_2$ in QE kinematics is naturally explained in
terms of scale separation in effective field theory.
This explanation has powerful consequences, allowing us to calculate and
predict SRC scaling factors from \textit{ab initio} low-energy nuclear
theory.
We present \textit{ab initio} calculations of SRC scaling factors
for a nucleus $A$ relative to the deuteron $a_2(A/d)$ and relative to
\isotope[3]{He} $a_2(A/\isotope[3]{He})$ in light and medium-mass nuclei.
Our framework further predicts that the EMC effect and SRC scaling
factors have minimal or negligible isovector corrections.\\

\noindent Keywords: short-range correlations, short-range-correlation scaling factors, EMC effect, quantum Monte Carlo
\end{abstract}

\section{Introduction}
\label{sec:intro}
The accurate description and prediction of the structure and behavior of
atomic nuclei remains an important problem in physics.
In spite of decades of experimental, theoretical, and computational
research and the fact that quantum chromodynamics (QCD) is widely
understood to provide the underlying field theoretic description, the
strong interaction between protons and neutrons still surprises us with
its subtlety.
Because of the nature of the nonabelian gauge interactions, QCD resists
perturbative treatments at low
energies~\cite{Gross:1973id,Politzer:1973fx}.
Explicit solutions at these energies are possible via the computational
framework of lattice QCD, wherein observables are calculated directly in
QCD but on a finite Euclidean space-time
lattice~\cite{Beane:2010em,Aoki:2012tk,Yamazaki:2015asa}.

While lattice QCD promises a fundamental explanation of nuclear physics
phenomena, the computational difficulties it faces grow rapidly with the
system size, so that current simulations are limited to few-nucleon
systems~\cite{Beane:2010em,Aoki:2012tk,Beane:2012vq,Yamazaki:2015asa,
Savage:2016kon,Wagman:2017tmp,Shanahan:2017bgi}.
This means that for many interesting nuclear systems, other methods are
needed at present.
Low-energy \textit{ab initio} nuclear theory, working with protons and
neutrons as degrees of freedom and fixing the parameters of the theory
with results from either experimental data or lattice QCD, naturally
fills this role, and the field has made significant progress in
recent years in terms of working with systematically improvable
Hamiltonians derived from chiral effective field theory (EFT) and in
terms of the size of the nuclear systems that can be accurately
handled~\cite{Barrett:2013nh,Soma:2013xha,Hagen:2013nca,Lahde:2013uqa,
Carlson:2014vla,Hergert:2015awm,Lynn:2017fxg,Lonardoni:2018nob}.

In particular, in recent years advances made in accurate quantum Monte
Carlo (QMC) methods and their combination with interactions derived
from chiral EFT has provided many new insights in low-energy nuclear
theory (see~\cite{Lynn:2019rdt} for a review).
One such insight to arise from the use of QMC methods with EFT
techniques is that, while commonly calculated two-body central densities
\begin{align}
	\rho_{2,\mathbbm{1}}(A,r)\propto \langle\Psi|\sum_{i<j}^A\delta(r-r_{ij})|\Psi\rangle\,,
\end{align}
with $r_{ij}$ the internucleon separation for a nucleus with $A$
nucleons, are scheme and scale dependent, their ratios are largely
scheme and scale independent~\cite{Chen:2016bde}.
Moreover, these ratios at small internucleon separation correspond to
short-range-correlation (SRC) observables in quasielastic (QE)
lepton-nucleus scattering: in short,
$\lim_{r\to0}\frac{\rho_{2,\mathbbm{1}}(A,r)}{\rho_{2,\mathbbm{1}}(d,r)}\propto
a_2(A/d)$, where $a_2$ is the so-called SRC scaling factor,
and $d$ stands for the deuteron~\cite{Chen:2016bde}.

In this paper, we exploit this unique convergence of advances in QMC
methods and EFT to confirm this relationship in light nuclei up to
\isotope[12]{C} by comparing with existing experimental data.
We then make predictions for several light systems (\isotope[6]{He},
\isotope[6]{Li}, and \isotope[16]{O}) and for the medium-mass nucleus
\isotope[40]{Ca}, which could be tested in existing and near-term future
experimental facilities.

The structure of this article is as follows.
In what remains of~\cref{sec:intro} we present some
background (\cref{subsec:back}), the main EFT arguments
(\cref{subsec:eft}), and details on the EFT power counting
(\cref{subsec:powcount}).
In~\cref{sec:hamqmc} we briefly discuss the nuclear Hamiltonian and our QMC
methods.
In~\cref{sec:res} we present our main results, discussing how best to
extract the SRC scaling factors from our QMC results.
Finally, in~\cref{sec:sum} we summarize our results and provide an
outlook for this novel framework.

\subsection{Background}
\label{subsec:back}
\begin{figure}[t]
	\centering
	\includegraphics[width=0.5\columnwidth]{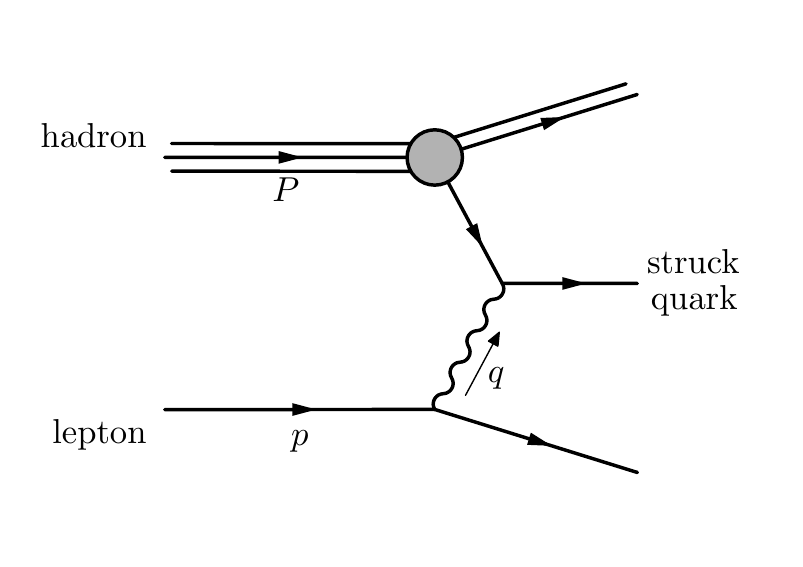}
	\caption[]{\label{fig:dis}
	Lowest-order DIS diagram. A highly energetic lepton of four momentum
	$p$ scatters from a struck quark inside a hadronic target of four
	momentum $P$, transferring four momentum $q$.}
\end{figure}

Deep-inelastic scattering (DIS) of leptons on nuclear targets has been
one of the most valuable experimental tools for learning about the structure
of nucleons and nuclei.
In DIS, a highly energetic ($Q^2\sim5\,\text{GeV}^2$) leptonic probe
with four momentum $p$ is scattered from a hadronic target with four
momentum $P$, transferring four momentum $q$ to the struck quark,
see~\cref{fig:dis}.
The cross section can be written in terms of the dimensionless
Bjorken $x\equiv\frac{Q^2}{2P\cdot q}$, with $Q^2=-q^2$, because $q$
is spacelike, the dimensionless variable $y\equiv\frac{P\cdot q}{P\cdot p}$,
and the structure function $F_2(x,Q^2)$:
\begin{align}
	\frac{{\rm d}^2{\sigma}}{{\rm d} x\,{\rm d} Q^2}=
	\frac{2\pi\alpha^2}{x\,Q^4}F_2(x,Q^2)\left[1+(1-y)^2\right]\,.
\end{align}

In 1983, the European Muon Collaboration (EMC) announced their
unexpected results for the measurement of the structure functions
$F_2^A(x,Q^2)$ in leptonic DIS for iron
(\isotope[56]{Fe}) and deuterium~\cite{Aubert:1983xm}.
The surprise came because, given that the typical binding energy per
nucleon is so small (i.e. $\lesssim1\%$) compared to the nucleon mass and
the energy transfer in the DIS process, the expectation was that the cross
section would have only trivial dependence on the nuclear target.
Instead, in the region $0.2\lesssim x\lesssim0.7$, the ratio
$2F_2^{\isotope{Fe}}/AF_2^{d}$ was observed to fall off linearly to a
significant reduction of $\sim10\%$ at $x\sim0.7$.
This reduction in the ratio
\begin{align}
R_\text{EMC}(A,x)\equiv\frac{2\,F_2^A(x,Q^2)}{A\,F_2^{d}(x,Q^2)}
\end{align}
has come to be known as the EMC effect.
Since then, significant experimental and theoretical
effort has been invested to understand this effect 
(see~\cite{Geesaman:1995yd,Norton:2003cb,Malace:2014uea,Hen:2016kwk}
for reviews).

\begin{figure}[t]
	\centering
	\includegraphics[width=0.495\columnwidth]{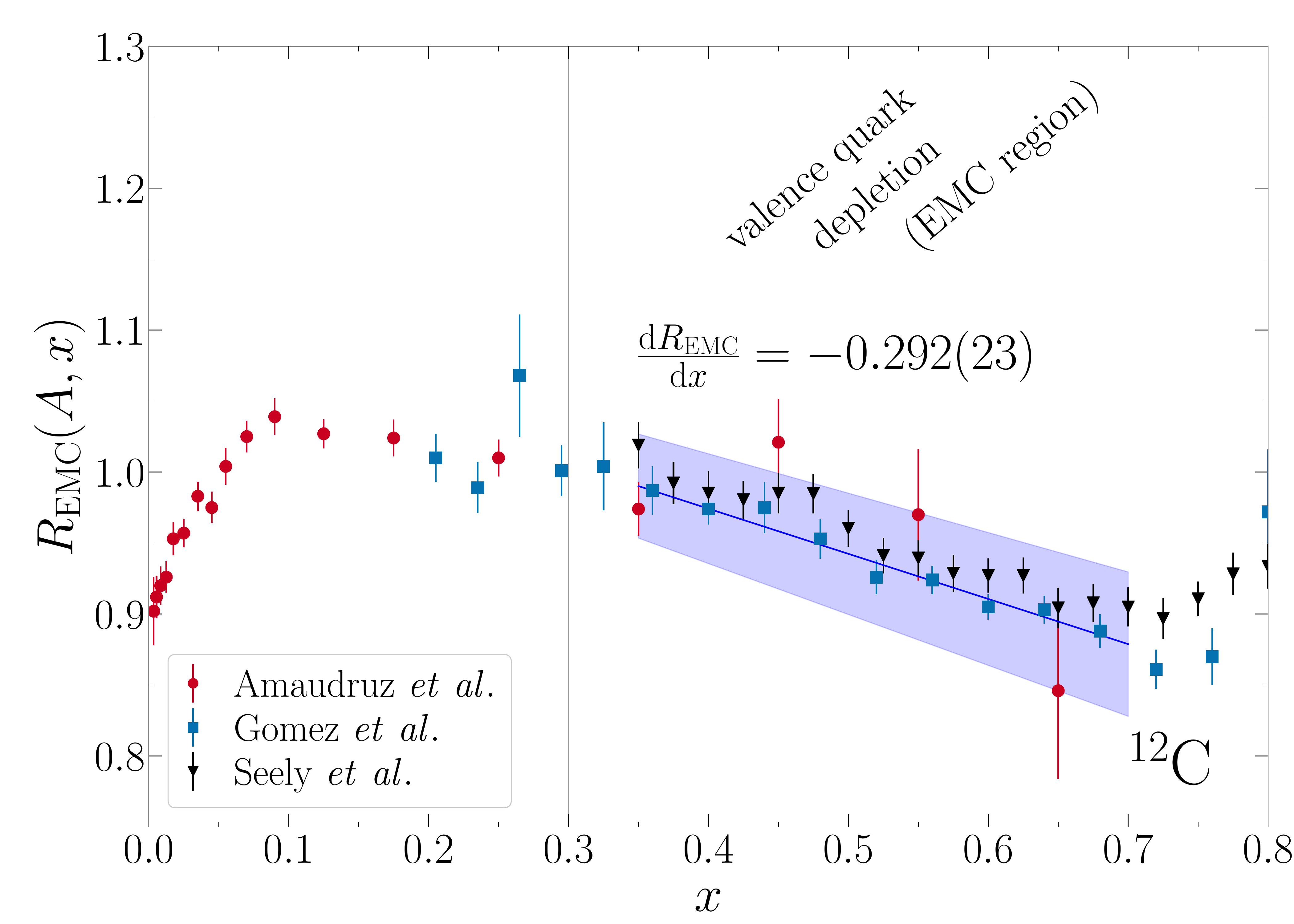}
	\includegraphics[width=0.495\columnwidth]{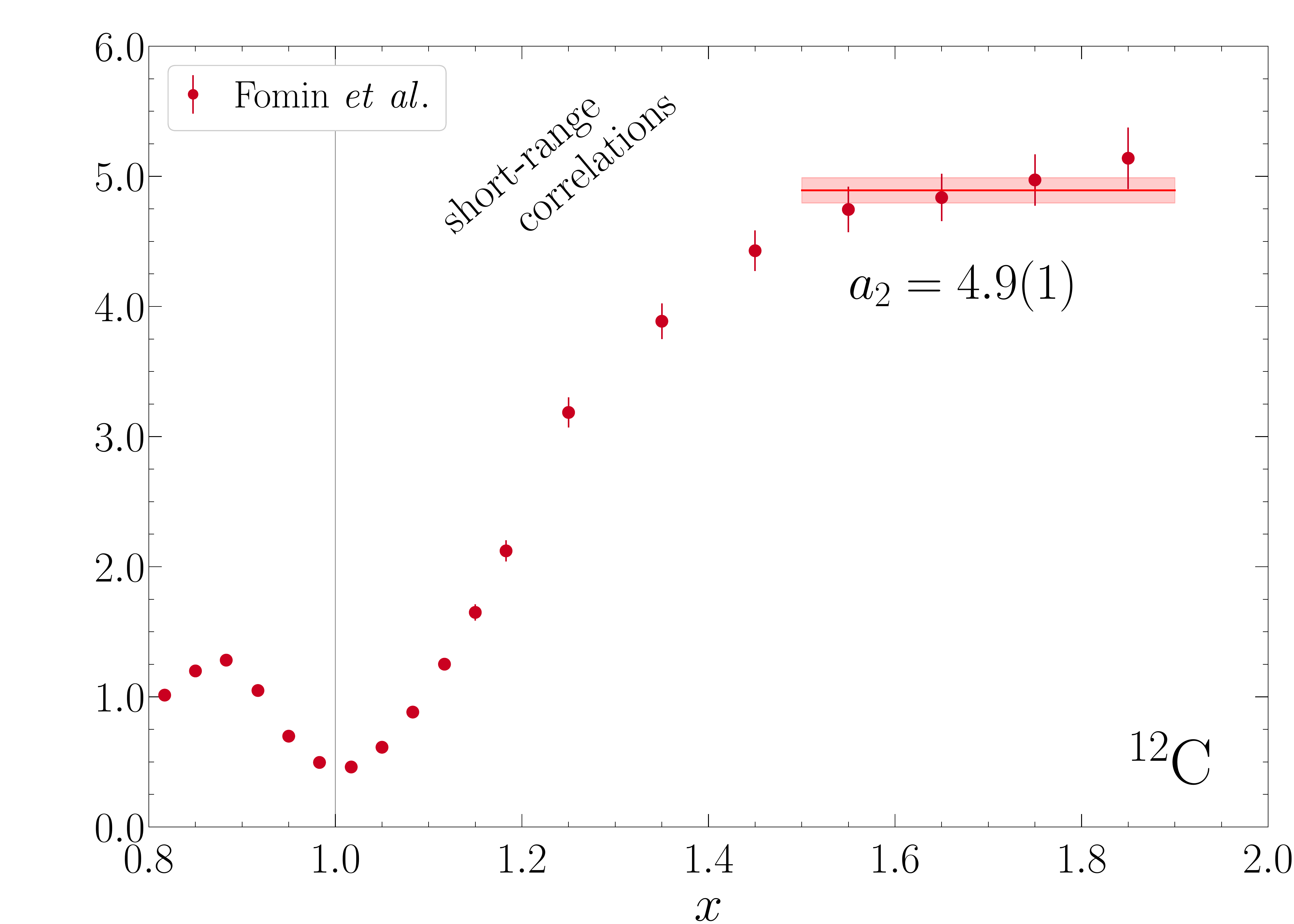}
	\caption[]{\label{fig:c12emc}
	An example of data for the ratio $R_\text{EMC}(A,x)$ collected in DIS
	(left panel with $0\le x\le0.8$) and QE experiments (right panel with
	$0.8\le x\le 2.0$) on \isotope[12]{C} showing the universal shape of the
	$x$ dependence of the EMC effect (note the different scales for the
	$y$ axes).
	The data are often separated into regions labeled for the favored
	explanation for the behavior in that region.
	Also shown are an illustrative linear fit to the EMC region of the
	Gomez~\textit{et al} data (left panel) and an illustrative fit to
	the plateau/SRC region (right panel).
	The data are from Amaudruz~\textit{et al}~\protect\cite{Amaudruz:1995tq},
	Gomez~\textit{et al}~\protect\cite{Gomez:1993ri}, and
	Seely~\textit{et al}~\protect\cite{Seely:2009gt} (left panel), and
	Fomin~\textit{et al}~\protect\cite{Fomin:2011ng,Fomin:2011ng.data} (right panel).
	}
\end{figure}

As part of this effort to further understand the implications of the EMC
effect, more experiments were carried out for smaller values of $x$, at
different $Q^2$~\cite{Amaudruz:1995tq}, and for various
nuclei~\cite{Gomez:1993ri},
and more recently in QE scattering at higher $x$, $1\lesssim x\lesssim 2$~\cite{Fomin:2011ng}.
The picture that emerges is that the ratio of nuclear structure
functions $R_\text{EMC}(A,x)$ has very little $Q^2$ dependence, and for
isoscalar nuclei, the $A$ and $x$ dependence of $R_\text{EMC}-1$
factorizes.
That is, the shape of the deviation from unity of the ratio
$R_\text{EMC}(A,x)$ is independent of $A$, while the maximum magnitude
only depends on $A$.
\Cref{fig:c12emc} shows an example of the universal $x$
dependence of the data.
The different regions are labeled with the favored explanation for the
behavior of the ratio in that region (see~\cite{Norton:2003cb} for
a more detailed explanation of the history of attempts at explaining the
EMC effect).

In this work, we are interested in the EMC region $(0.35<x<0.7)$ and the
SRC region $(1<x<2)$.
The strength of the effect in the former region is usually characterized
by the slope $|{\rm d} R_\text{EMC}/{\rm d} x|$ (see
again~\cref{fig:c12emc}), which ranges from $\sim0.07$ in
\isotope[3]{He} up to $\sim0.5$ in \isotope[108]{Ag}, showing a trend
towards saturation as the mass number $A$ increases.
In the latter region, based on an impulse-approximation argument,
Frankfurt~\textit{et al}~\cite{Frankfurt:1993sp} cast the inclusive
cross section as
\begin{align}
	\sigma(x,Q^2)=\sum_{j=2}^A\frac{1}{j}\,a_j(A)\,\sigma_j(x,Q^2)\, ,
\end{align}
where the $a_j(A)$ are proportional to the probabilities to find a nucleon in
a $j$-nucleon SRC, and $\sigma_j(x,Q^2)=0$ for $x>j$.
This framework correctly predicted the scaling behavior ($x$ and
$Q^2$ independence) in the ratio of cross sections:
\begin{align}
	a_2(A/d)\equiv\left.\frac{2\sigma_A}{A\sigma_d}\right|_{1.5<x<2}\,,
\end{align}
where $a_2$ is the SRC scaling factor introduced earlier.
(Note that Fermi motion pushes the onset of the plateau from $x\sim1$ to
$x\sim1.5$).
In some of the more recent experiments at Thomas Jefferson National
Accelerator Facility (Jefferson Lab)~\cite{Fomin:2011ng}, these plateaus
have been observed for nuclear targets from \isotope[3]{He} to
\isotope[197]{Au} (see also~\cref{fig:c12emc}).

\begin{figure}[t]
	\centering
	\includegraphics[width=0.5\columnwidth]{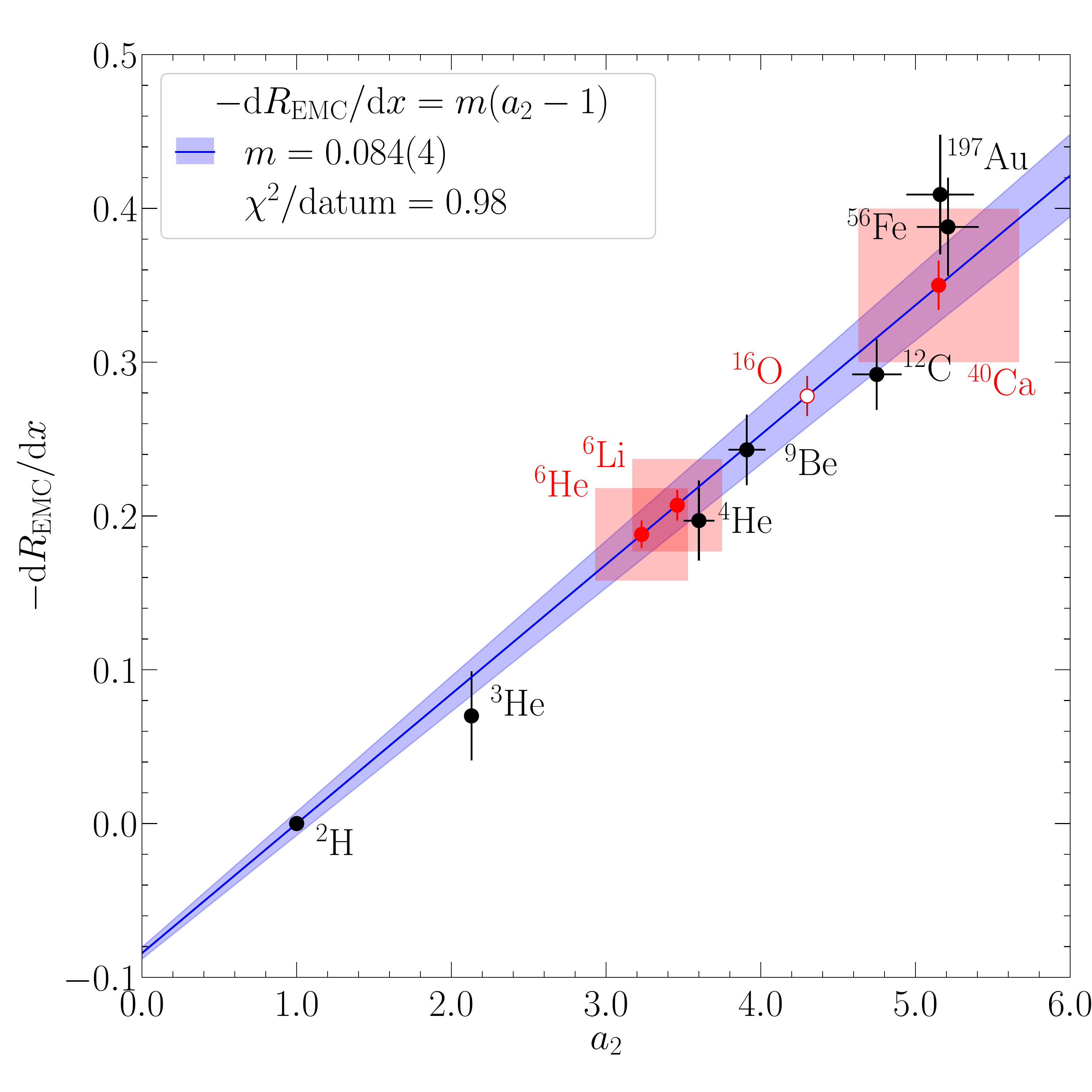}
	\caption[]{\label{fig:drdx_a2}
	The linear relationship between the strength (slope) of the EMC
	effect $-{\rm d} R_\text{EMC}/{\rm d} x$ and the SRC scaling factor $a_2$.
	The fit is constrained to pass through the deuteron point with
	$-{\rm d} R_\text{EMC}/{\rm d} x = 0$, and $a_2=1$: hence the form of the fit
	$-{\rm d} R_\text{EMC}/{\rm d} x=m(a_2-1)$. Data (in black) are taken from~\protect\cite{Hen:2012fm}.
	In red are our predictions from this work for \isotope[6]{He},
	\isotope[6]{Li}, and \isotope[16]{O} using local chiral EFT
	interactions at \nxlo{2} with the $E\tau$ parametrization of the
	$3N$ interaction, and for \isotope[40]{Ca} using the simplified
	$\text{AV}4'+\text{UIX}_\text{c}$ potential (see~\cref{tab:a2}
	and~\cref{sec:res}).
	The QMC statistical uncertainties are shown as the red error bars
	(the horizontal statistical uncertainties are smaller than the
	points).
	The systematic errors coming from the truncation of the chiral
	expansion (where available) and from the fit of $a_2$ are shown as
	the red shaded areas.
	For \isotope[16]{O} (the empty red circle), we do not show the
	associated systematic uncertainties as they are large enough
	(see~\cref{tab:a2}) as to obscure the figure.
	}
\end{figure}

Recently, a fascinating empirical discovery was made: the slope of the
EMC effect in the EMC region is linearly correlated with the SRC scaling
factor~\cite{Weinstein:2010rt,Hen:2012fm}, see~\cref{fig:drdx_a2}.
This remarkable result has motivated a series of experiments attempting
to further understand this phenomenon, as well as many theoretical
proposals.
As discussed in~\cite{Chen:2016bde}, the physics behind this
correlation is naturally explained in the EFT approach used here.

\subsection{Effective field theory}
\label{subsec:eft}
Effective field theory is a model-independent approach that
relies on the symmetries and the separation of scales in a given
system.
Effective field theory has been successfully applied to many aspects of
meson~\cite{Gasser:1983yg}, single-~\cite{Bernard:1995dp}, and
multi-nucleon
systems~\cite{beane2012,beane2001,bedaque2002,kubodera2004,epelbaum2009,hammer2013}.
In particular, chiral EFT has been applied to parton distribution
functions (PDFs) in the meson,
single-nucleon~\cite{AS,CJ1,CJ2,DMNRT,DMT1,DMT2,Detmold:2005pt,Hagler:2007xi,Gockeler:2003jfa},
and multi-nucleon sectors~\cite{Chen:2004zx,Beane:2004xf}, as well as to
other light-cone dominated
observables~\cite{CJ3,Belitsky:2002jp,Chen:2003fp,Chen:2006gg,
Ando:2006sk,Diehl:2006ya}.

In 2005, using EFT, Chen and Detmold~\cite{Chen:2004zx} found that, up
to higher order corrections, the $F_2$ structure function of an
isoscalar nucleus has the form
\begin{align}
	\label{eq:factorizationX}
	F_2^A(x,Q^2)/A\simeq F_2^N(x,Q^2)+g_2(A,\Lambda)f_2(x,Q^2,\Lambda)\, ,
\end{align}
where $F_2^N$ is the isoscalar combination of the nucleon structure
function, which receives the nuclear modification from the second term
in which the $x$ and $A$ dependence factorizes.
The $A$ dependence comes from momenta smaller than the ultraviolet
momentum cutoff of the EFT $\Lambda\sim0.5\,\rm GeV$, while the $x$
dependence comes from scales larger than $\Lambda$. 

An immediate consequence of \cref{eq:factorizationX} is that 
\begin{align}
	\label{eq:REMC}
	R_\mathrm{EMC}(A,x)-1
	\simeq
	C(x)\left[a_{2}(A)-1 \right]\, ,
\end{align}
with the $x$ and $A$ dependence factorized, and 
\begin{align}
	\label{eq:Cofx}
	C(x)&=1-\frac{2F_2^N(x)}{F_2^d(x)}\, ,\\
	\label{eq:a2}
	a_{2}(A)&=\frac{g_{2}(A,\Lambda )}{g_{2}(2,\Lambda)}\, .
\end{align}
The deviation of $R_\text{EMC}$ from unity in \cref{eq:REMC} means that
the nuclear modification to the structure functions has a universal
shape ($x$ dependence), while its maximum magnitude depends only on
$A$~\cite{Chen:2004zx}.
This feature describes experimental data with $x<1$ for many nuclei,
ranging from \isotope{He} to \isotope{Pb} very
well~\cite{Frankfurt:1988nt,Frankfurt:1981mk}.  

Because $F_2^A(x)$ has support for $0<x<A$, if DIS experiments were
carried out at $1<x<2$, where $F_2^N(x)=0$ but $F_2^d(x)\ne0$, then
\cref{eq:REMC,eq:Cofx} yields 
\begin{align} 
	\label{eq:a2x}
	R_\text{EMC}(A,1<x<2)\simeq a_{2}(A)\, ,
\end{align}
which is an $x$-independent plateau.
Experimentally, the measurements at $x>1$ are performed not in the DIS
region, but in the QE region at lower $Q^2$ because of the larger
associated rate.
Generalizing the analysis to the QE region by including all the higher
twist effects does not change the plateau value
of~\cref{eq:a2x}~\cite{Chen:2016bde}.
The plateau is observed experimentally at $1.5<x<2$, possibly because
Fermi motion, which is a higher-order effect in the EFT, extends the
contribution of the single-nucleon PDF to $x$ slightly above 1, so that
the onset of the plateau is also pushed to larger $x$.

From~\cref{eq:REMC,eq:a2x}, the observed linear relation between
$-{\rm d} R_\text{EMC}/{\rm d} x$ and the SRC scaling factor $a_{2}(A)$ is
easily obtained.
\Cref{eq:a2x} demands that the scaling factor, which comes from the ratio
of two cross sections, be independent of the cutoff $\Lambda$. 
Therefore, the $\Lambda$ dependence on the right-hand side of~\cref{eq:a2}
should cancel. 
This provides a nontrivial test of EFT, because it implies that, although
$g_2(A,\Lambda)$ depends on the renormalization scheme and scale ($\Lambda$) of the
EFT, $a_2$ is scheme and scale independent.
This occurs if the $\Lambda$ and $A$ dependence factorize in $g_2$,
which is defined as
\begin{align}
	g_2(A,\Lambda)\equiv\frac{1}{2A}
	\langle A|\normord{(N^\dagger N)^2}|A\rangle_\Lambda\, ,
\end{align}
where $N$ is the nucleon field and $\normord{\ \cdots\ }$ indicates
normal ordering of the enclosed operators with respect to the vacuum
state.

The above analysis is for isoscalar operators.
Including isovector corrections, one has
\begin{align}
	\label{eq:factorizationXX}
	\begin{split}
	F_2^A(x,Q^2)&\simeq ZF_2^p(x,Q^2)+NF_2^n(x,Q^2)+Ag_2(A,\Lambda)f_2(x,Q^2,\Lambda)+\cdots\, ,
	\end{split}
\end{align}
with $N$ $(Z)$ the number of neutrons (protons) in the nucleus.
The isovector counterpart of the $g_2$ term is neglected because it is
$\mathcal{O}((N-Z)/A N_c)$ smaller than $g_2$, with the number of
colors $N_c=3$.
This implies that, even with isovector corrections, the SRC plateaus
still exist, and the plateau values of $a_2$ remain unchanged.
Also, for the EMC effect, recent experimental results including
nonisoscalar nuclei are well described
by~\cref{eq:factorizationXX}~\cite{Schmookler:2019nvf}.

\subsection{EFT power counting}
\label{subsec:powcount}
In DIS, the structure functions $F_{2}^{A}(x,Q^2)$ can be expressed in
terms of nuclear PDFs $q_{i}^{A}(x,Q)$ as
$F_{2}^{A}(x,Q^2)=\sum_{i}Q_{i}^{2}xq_{i}^{A}(x,Q)$, where the sum
runs over quarks and antiquarks of flavor $i$ with charge $\pm Q_i$ in a
nucleus $A$.
In what follows, we first focus on the power counting for
isoscalar PDFs, $q_A=q_{A,0}=q_{A,u}+q_{A,d}$, then we discuss the
isovector correction from $q_{A,3}=q_{A,u}-q_{A,d}$.
The dominant (leading-twist) PDFs are determined by target matrix
elements of bilocal light-cone operators.
Applying the operator product expansion, the Mellin moments of the
PDFs,
\begin{align}
	\langle x^n\rangle_A(Q)=\int_{-A}^{A}{\rm d} x\,x^nq_A(x,Q)\,, 
\end{align}
are determined by matrix elements of local operators,
\begin{align}
	\label{eq:melo}
	\langle A;p|\mathcal{O}^{\mu_{0}\cdots\mu_{n}}|A;p\rangle=
	2\langle x^n\rangle_A(Q)\,p^{(\mu_0}\cdots p^{\mu_n)}\,,
\end{align}
with
\begin{align}
	\label{eq:O}
	\mathcal{O}^{\mu_{0}\cdots\mu_{n}}=
	\overline{q}\gamma^{(\mu_{0}}\ii D^{\mu_{1}}\cdots\ii D^{\mu_{n})}q\,,
\end{align}
where $(\cdots)$ indicates that the enclosed indices have been symmetrized and
made traceless, $D^\mu\equiv(\vec{D}^\mu-\cev{D}^\mu)/2$ is the covariant
derivative, and a sum over flavors $q=u,d$ is implied.
The negative $x$ distribution is the antiquark distribution:
$q_A(-x)=-\bar{q}_A(x)$. 

In nuclear matrix elements of these operators, there are other relevant
momentum scales below the hard scattering scale $Q$:
$\Lambda\sim 0.5\,\rm GeV$ is the range of validity of the EFT, and
$P\sim m_{\pi}$ is a typical momentum inside the nucleus ($m_\pi$ is the
pion mass).
These scales satisfy $Q\gg\Lambda\gg P$, and the ratio $\Lambda/Q$ is the
small expansion parameter in the twist expansion, while the ratio
$\epsilon\sim P/\Lambda\sim 0.2\text{--}0.3$ is the small expansion parameter
for the chiral expansion.

In EFT, each of the QCD operators is matched to a sum of all possible
hadronic operators of the same symmetries at the scale
$\Lambda$~\cite{Chen:2004zx}
\begin{align}
	\begin{split}
	\label{eq:matching}
	\mathcal{O}^{\mu_{0}\cdots\mu_{n}}\to 
	&:2\langle x^n \rangle_{N}M_N^{n+1}v^{(\mu_{0}}\cdots v^{\mu_{n})}
	N^\dagger N\left[1+\alpha_{n}N^\dagger N\right]\\
	&+\langle x^n \rangle_{\pi}\pi^\alpha\ii\partial^{(\mu_{0}}\cdots\ii\partial^{\mu_{n})}
	\pi^\alpha+\cdots\!:\,,  
	\end{split}
\end{align}
where $\pi$ ($N$) is the pion (nucleon) field, $v$ is the nucleon four-velocity, and
$\langle x^n \rangle_{N\,(\pi)}$ is the $n$th moment of the isoscalar quark PDF in a free
nucleon (pion).
There are an infinite number of terms on the right-hand side
of~\cref{eq:matching}, whose importance will be estimated by power
counting.
The $\langle x^n \rangle_{N\,(\pi)}$ terms are one-body operators acting on a
single hadron, whose prefactors can be determined by taking the
nucleon (pion) matrix element of~\cref{eq:matching}.
The $\alpha_n$ terms are two-body operators.
Here we have only kept the SU(4) (spin and isospin) singlet two-body
operator $\propto\left(N^\dagger N\right)^2$ and neglected the SU(4)
nonsinglet operator $\propto(N^\dagger\bm\sigma N)^2-
(N^\dagger\bm\tau N)^{2}$, which changes sign when interchanging the spin
($\bm\sigma$) and isospin ($\bm\tau$) matrices~\cite{Mehen:1999qs}.
The latter operator has an additional $\mathcal{O}(1/N_{c}^{2})\sim0.1$
suppression in its prefactor \cite{Kaplan:1995yg}.
We also replace the nucleon velocity by the nucleus velocity and include
the correction $\ii\partial_0/M_N$ at higher orders.

In Weinberg's power counting scheme, the typical nucleon momenta
$|{\bf q}|$ are counted as $\mathcal{O}(\epsilon)$, while their energies
$q^0$ are $\mathcal{O}(\epsilon^2)$.
Two-nucleon contact operators $(N^\dagger N)^2$ are counted as
$\mathcal{O}(\epsilon^0)$, while the three-body contact operator
$(N^\dagger N)^3$ is counted as $\mathcal{O}(\epsilon^3)$, both
according to their mass dimension. 
We will focus on the twist-2 operators with all $\mu_i=0$
in~\cref{eq:matching}.
Because $v^0=1$, the $v^{(\mu_0}\cdots v^{\mu_n)}(N^\dagger N)$ operator
is $\mathcal{O}(\epsilon^{-3})$ and $v^{(\mu_0}\cdots
v^{\mu_n)}(N^\dagger N)^2$ is $\mathcal{O}(\epsilon^0)$.
The one-derivative operator $N^\dagger\partial^{(\mu_0}v^{\mu_1}\cdots
v^{\mu_n)}N$ is $\mathcal{O}(\epsilon^{-1})$, but its net
effect is to shift the value of $p^0$ on the right-hand side
of~\cref{eq:melo} from $AM_N$ to $M_A$.
This can be seen from the special case of $n=0$.
The vector current operator $\mathcal{O}^{\mu_0}$ is matched to the
operator $2M_NN^\dagger(v^{\mu_0}+\ii\partial^{\mu_0}/M_A)N$.
The nuclear matrix element of the first term yields $2AM_N$.
The relative coefficient between the two terms are fixed by
reparametrization invariance~\cite{Luke:1992cs}, and the nuclear matrix
element of the sum yields $2M_A$.

The two-derivative operator given by
$N^\dagger\partial^{(\mu_0}\partial^{\mu_1}v^{\mu_2}\cdots v^{\mu_n)}N$,
(again with $\mu_i=0$ for all $i$) is $\mathcal{O}(\epsilon)$, and it
can cause $q_N(x)$ or $F^N_2(x)$ to ``spill'' into $x>1$.
This is related to Fermi motion.
Although it is higher order than the two-body operator, if $f_2(x)$
of~\cref{eq:factorizationX} is very small when $x$ is just above one,
then the Fermi-motion effect could be dominant and explain why the
$a_2$ plateau only sets in at $x\gtrsim 1.5$.
It is important to note that, in the EFT approach, off-shell effects that
enter through Fermi motion can be absorbed into many-body operators through
a field redefinition~\cite{Ji:2013dva,Lin:2014zya}.
Therefore the separation between ``Fermi motion'' and ``two-body effects''
is meaningful only after the theory is clearly specified.

The pion one-body operator $\pi^{a}\ii\partial^{(\mu_0}\cdots
\ii\partial^{\mu_n)}\pi^{a}$ inserted in the one-pion-exchange diagram
contributes at $\mathcal{O}(\epsilon^{n-1})$.
Because $\langle x^n \rangle_\pi=0$ for even $n$ due to charge conjugation symmetry,
the $n=1$ pion operator enters at $\mathcal{O}(\epsilon^0)$, but for
higher $n$ the contributions either vanish or are higher order compared
with the other operators in~\cref{eq:matching}.
This means that, at $\mathcal{O}(\epsilon^0)$, the pion contribution
to~\cref{eq:factorizationX} is proportional to $\delta(x)/x$ and breaks the
factorization of the $x$ and $A$ dependence of $F^2_A$
of~\cref{eq:factorizationX}, but only at $x=0$. 

All the other operators in the matching are found to be higher order than
$\epsilon^0$ in this power counting.
Using nucleon number conservation, $\langle A|\normord{N^\dagger N}|A\rangle=A$, the
isoscalar nuclear matrix element of~\cref{eq:matching} is
\begin{align}
	\label{eq:xA}
	\begin{split}
	\langle x^n \rangle_A(Q)&=\langle x^n \rangle_{N}(Q)\Bigl[A+\alpha_n(\Lambda,Q)
	\langle A|\normord{(N^\dagger N)^2}|A\rangle_\Lambda\Bigr]\, \\
	&+\delta_{n=1}\text{ term} \, ,
	\end{split}
\end{align}
where $\alpha_n$ is $A$ independent but $\Lambda$ dependent, and is
completely determined by the two-nucleon system.
After an inverse Mellin transform, except at $x=0$ as explained above,
the isoscalar PDFs satisfy
\begin{align}
	\label{eq:qA}
	q_A(x,Q)/A\simeq q_N(x,Q)+g_2(A,\Lambda)\tilde{q}_2(x,Q,\Lambda)\,,
\end{align}
where $\tilde{q}_2(x,Q,\Lambda)$ is an unknown function independent of
$A$ whose Mellin moments are determined by the low-energy constants
$\alpha_n$.
This result also holds at the level of the structure function,
which leads to~\cref{eq:factorizationX}.

The isovector operator 
\begin{align}
	\label{eq:isoO}
	\mathcal{O}_3^{\mu_0\cdots\mu_n}=\overline{q}\tau_3\gamma^{(\mu_0}\ii
	D^{\mu_1}\cdots\ii D^{\mu_n)}q\,,
\end{align}
is matched to hadronic operators as
\begin{align}
	\label{eq:matchingX}
	\begin{split}
	\mathcal{O}_3^{\mu_0\cdots\mu_n}\to
	&:2\langle x^n \rangle_{N,3}M_N^{n+1}
	v^{(\mu_0}\cdots v^{\mu_n)}N^\dagger\tau_3N\left[1+\gamma_nN^\dagger N\right]\\
	&+2\delta_nM_N^{n+1}N^\dagger S^{(\mu_0}v^{\mu_1}\cdots v^{\mu_n)}
	\pi^{\alpha}[\tau^{\alpha}, \tau_3]N\\
	&+\langle x^n \rangle_{\pi,3}\ii\epsilon^{3\alpha\beta}\pi^\alpha\ii
	\partial^{(\mu_0}\cdots\ii\partial^{\mu_n)}\pi^{\beta}+\cdots\!:\,.  
	\end{split}
\end{align}
The $\langle x^n \rangle_{N,3}$ term is $\mathcal{O}(\epsilon^{-3})$.
The $\gamma_n$ term is $\mathcal{O}(\epsilon^0)$, like the $\alpha_n$
operator of~\cref{eq:matching}, but it has an additional $1/N_c$
suppression in its prefactor~\cite{Kaplan:1995yg} and an $(N-Z)/A$
suppression in its nuclear matrix element compared with the $\alpha_n$
term and, hence can be neglected.
$S^{\mu}$ is the nucleon spin vector. 
Using $\pi^\alpha[\tau^\alpha,\tau_3]\propto(\pi^+\tau_+-\pi^-\tau_-)$, 
the $\delta_n$ term involves a charged pion exchange, which can only happen
between $np$ states in two-nucleon systems.
However, $\tau_3$ for $np$ states (which have isospin zero) vanishes,
therefore, there is no net two-nucleon contribution from this term.
The $\langle x^n \rangle_{\pi,3}$ term contributes at $\mathcal{O}(\epsilon^{n-1})$.
However, $\langle x^n \rangle_{\pi,3}$ vanishes for odd $n$ by charge conjugation.
The $\langle x^0\rangle_{\pi,3}$ term is the isospin charge, which is protected from
nuclear modifications.
The other terms $\langle x^{n\ge2}\rangle_{\pi,3}$ are $\mathcal{O}(\epsilon)$ and
higher and can be neglected.
The leading three-body operator $v^{(\mu_0}\cdots
v^{\mu_n)}(N^\dagger N)^2N^\dagger\tau_3N$ is $\mathcal{O}(\epsilon^3)$,
which can also be neglected.

We remark that in the large $N_c$ limit, the nucleon and delta resonances are degenerate, 
hence one should explicitly include the deltas in the $1/N_c$ expansion. In the real world with $N_c=3$, 
the mass difference between delta and nucleon $\Delta m$ is much larger than the typical Fermi energy $E_F$ 
in a nucleus. Therefore, one can choose to integrate out the delta degrees of freedom, 
as done in this work. The effect is that the nucleon operators studied here will receive 
$\mathcal{O}(E_F/\Delta m)$ corrections, but their $N_c$ scalings remain the same.

In summary, up to $\mathcal{O}(\epsilon^0)$, only the one-body operator
$\langle x^n \rangle_{N,3}$ contributes to isovector corrections.
Therefore, the nuclear effects are dominated  by the isoscalar PDF
contributions, while the isovector PDFs are relatively unaltered by the
nuclear environment, leading to~\cref{eq:factorizationXX}.

\section{Hamiltonian and Quantum Monte Carlo Methods}
\label{sec:hamqmc}
In \textit{ab initio} methods, nuclei are treated as a collection of
point-like particles of mass $M_N$ interacting via two- and three-body
potentials according to the nonrelativistic Hamiltonian
\begin{align}
	\label{eq:ham}
	H=-\frac{\hbar^2}{2M_N}\sum_i^A\nabla^2_i+\sum_{i<j}^Av_{ij}+
	\sum_{i<j<k}^AV_{ijk}\,,
\end{align}
where the two-body interaction $v_{ij}$ also includes the Coulomb force.

In this work, we adopt the local chiral nucleon-nucleon ($N\!N$) interactions
at next-to-next-to-leading order (\nxlo{2}) in Weinberg counting of~\cite{Gezerlis:2013ipa,Gezerlis:2014zia},
with coordinate-space cutoffs $R_0=1.0\,\rm fm$ and $R_0=1.2\,\rm fm$.  
Such interactions include long-range pion-exchange contributions,
determined by pion-nucleon couplings, and shorter-range contributions,
defined by low-energy couplings (LECs) that are fit to reproduce $N\!N$ scattering
data.
The local chiral $N\!N$ potentials are written in coordinate space as a sum of
spin/isospin operators
\begin{align}
	\label{eq:v_ij}
	v_{ij}=\sum_{p=1}^7v_p(r_{ij})\mathcal{O}_{ij}^p\,,
\end{align}
with
\begin{align}
	\mathcal{O}_{ij}^{p=1,\ldots,7}=\Big\{&\mathbbm{1},\bm\tau_{i}\cdot\bm\tau_{j},
	\bm\sigma_{i}\cdot\bm\sigma_{j},\bm\sigma_{i}\cdot\bm\sigma_{j}\,\bm\tau_{i}\cdot\bm\tau_{j},
	S_{ij},S_{ij}\,\bm\tau_{i}\cdot\bm\tau_{j},{\bf L}\cdot{\bf S}\Big\}\,,
\end{align}
where $r_{ij}=|{\bf r}_i-{\bf r}_j|$ is the $N\!N$ relative distance, 
$S_{ij}=3\,\bm\sigma_i\cdot\hat{{\bf r}}_{ij}\,
\bm\sigma_j\cdot\hat{{\bf r}}_{ij}-\bm\sigma_{i}\cdot\bm\sigma_{j}$ 
is the tensor operator, and
${\bf L}=({\bf r}_i-{\bf r}_j)\times(\bm\nabla_i-\bm\nabla_j)/2\ii$ and
${\bf S}=(\bm\sigma_i+\bm\sigma_j)/2$ are the relative angular
momentum and the total spin of the pair $ij$, respectively.

At \nxlo{2}, in addition to the $N\!N$ interactions specified above, 
three-nucleon ($3N$) interactions 
enter~\cite{Tews:2015ufa,Lynn:2015jua,Lynn:2017fxg,Lonardoni:2018nob},
see also~\cite{Vankolck:1994,Epelbaum:2002} for earlier formulations, 
often used as non-local interactions in momentum space.
The employed $3N$ forces include two-pion-exchange (TPE) contributions 
in $P$ and $S$ waves, plus shorter-range components parametrized by two
contact terms, usually referred to as $V_D$ and $V_E$:
\begin{align}
	\label{eq:v_ijk}
	V_{ijk}=V_{2\pi}^{P}+V_{2\pi}^{S}+V_D+V_E\,.
\end{align}
The TPE components are characterized by the LECs $c_1$, $c_3$, and
$c_4$ from the pion-nucleon sector.
The LECs of the contact terms, $c_D$ and $c_E$, have been fit to the
$\alpha$ particle binding energy and to the spin-orbit splitting in the
neutron-$\alpha$ $P$-wave phase shifts~\cite{Lynn:2015jua,Lonardoni:2018nob}.
We employ the form
\begin{align}
	V_{D2}=\frac{g_Ac_Dm_\pi^2}{96\pi\Lambda_\chi
	F_\pi^4}\sum_{i<j<k}\sum_\text{cyc}\bm\tau_{i}\cdot\bm\tau_{k}\left[X_{ik}({\bf r}_{ik})-
	\frac{4\pi}{m_\pi^2}\bm\sigma_{i}\cdot\bm\sigma_{k}\delta_{R_{3N}}(r_{ik})\right]
	[\delta_{R_{3N}}(r_{ij})+\delta_{R_{3N}}(r_{kj})]\,,
\end{align}
for $V_D$, and we consider two choices for $V_E$, namely $E\tau$ and
$E\mathbbm{1}$:
\begin{subequations}
	\begin{align}
		V_{E\tau}&=\frac{c_E}{\Lambda_\chi F_\pi^4}
		\sum_{i<j<k}\sum_\text{cyc}\bm\tau_{i}\cdot\bm\tau_{k}\,\delta_{R_{3N}}(r_{kj})\,\delta_{R_{3N}}(r_{ij})\,,\\
		V_{E\mathbbm{1}}&=\frac{c_E}{\Lambda_\chi F_\pi^4}
		\sum_{i<j<k}\sum_\text{cyc}\,\delta_{R_{3N}}(r_{kj})\,\delta_{R_{3N}}(r_{ij})\,,
	\end{align}
\end{subequations}
where $g_A$ is the axial vector coupling constant, $m_\pi$ is the pion
mass, $\Lambda_\chi=700\,\text{MeV}$, $F_\pi$ is the pion decay constant,
$X_{ij}({\bf r}_{ij})=[S_{ij}({\bf r}_{ij})T(r_{ij})+\bm\sigma_{i}\cdot\bm\sigma_{j}]Y(r_{ij})$ is the
coordinate-space pion propagator, with the tensor and Yukawa functions
defined as $T(r)=1+3/(m_\pi r)+3/(m_\pi r)^2$ and $Y(r)=\eu{-m_\pi r}/r$,
respectively, and
$\delta_{R_{3N}}(r)=\frac{\eu{-(r/R_{3N})^4}}{\pi\Gamma(3/4)R_{3N}^3}$ is a
smeared-out delta function with $3N$ coordinate-space cutoff $R_{3N}$.
We take this $3N$ cutoff equal to the $N\!N$ cutoff $R_{3N}=R_0$.
The notation $\sum_\text{cyc}$ indicates a cyclic summation over the
indices $\{ijk\}$.
See~\cite{Lynn:2015jua,Lynn:2017fxg,Lonardoni:2018nob} for more details
including values for $c_D$ and $c_E$.

The operator structure of the employed local chiral interactions is
suited for QMC calculations.
QMC methods are a family of \textit{ab initio} many-body
techniques that allow one to solve the many-body Schr\"odinger equation
in a nonperturbative fashion with high accuracy.
In particular, imaginary-time projection algorithms, also known as
diffusion Monte Carlo (DMC) algorithms, have proven to be remarkably
successful in the description of nuclei and their global properties,
e.g. binding energies, radii, transitions, and reactions, and in the
prediction of properties of neutron star matter (for a review of QMC
methods see~\cite{Carlson:2014vla}).

In this work, we employ two different DMC techniques, namely the Green's
function Monte Carlo (GFMC) method~\cite{Carlson:1987zz} and the auxiliary
field diffusion Monte Carlo (AFDMC) method~\cite{Schmidt:1999lik}. 
Both approaches rely on the application of an imaginary-time propagator
to an initial trial wave function in order to project out the true
many-body ground state of the system:
\begin{subequations}
	\begin{align}
		|\Psi(\tau)\rangle&\equiv\eu{-H\tau}|\Psi_T\rangle\,,\\
		\lim_{\tau\to\infty}|\Psi(\tau)\rangle&\to|\Psi_0\rangle\,.
	\end{align}
\end{subequations}
The trial wave function is given in terms of a variational state of the
form
\begin{align}
	\label{eq:psi}
	|\Psi_T\rangle=\big[F_C+F_2+F_3\big]|\Phi\rangle_{J^\pi T}\,,
\end{align}
where $F_C$ accounts for all of the spin/isospin-independent correlations,
and $F_2$ and $F_3$ are spin/isospin-dependent two- and three-body
correlations, respectively.
The term $|\Phi\rangle$ is taken to be a shell-model-like state with total
angular momentum $J$, parity $\pi$, and total isospin $T$.
Its wave function is constructed using single-particle orbitals that
depend on the nucleon spatial coordinates, spin, and isospin. 
An initial optimization procedure is applied to the trial state
of~\cref{eq:psi} in order to find the optimal parameters providing the
best, i.e. lowest, variational energy. 
The optimized wave function is then repetitively evolved in small
imaginary-time steps until the ground state of the system is reached
(more details can be found in~\cite{Carlson:2014vla,Lonardoni:2018nob}).

The local chiral interactions considered in this work can be efficiently
implemented in both the GFMC and AFDMC methods.
The GFMC method, which includes a sum over all possible spin/isospin
states at each step in the diffusion, scales exponentially with the
number of nucleons $A$.
This limits current calculations to around $A=12$.
The AFDMC method, on the other hand, samples the sum over all
spin/isospin states, and therefore exhibits a much gentler, polynomial
scaling with $A$.
The two algorithms are thus complementary, and they allow one to vastly
extend the region of applicability of QMC calculations.
Results employing local chiral forces are now available for several
quantities (binding energies, charge radii, charge form factors, single-
and two-nucleon radial distributions, and single- and two-nucleon
momentum distributions) in light and medium-mass
nuclei~\cite{Lynn:2015jua,Lynn:2017fxg,Lonardoni:2017hgs,Lonardoni:2018nob,Lonardoni:2018sqo},
and for properties of pure neutron
systems~\cite{Zhao:2016ujh,Klos:2016fdb,Gandolfi:2016bth}, including pure neutron
matter~\cite{Gezerlis:2013ipa,Gezerlis:2014zia,Tews:2015ufa,Lynn:2015jua}. 

In QMC methods, the expectation value of an observable $\mathcal O$ is
calculated as
\begin{align}
	\label{eq:obs}
	\langle\mathcal{O}\rangle=\frac{1}{\mathcal{N}}\sum_{i=1}^\mathcal{N}
	\frac{\langle R_iS_i|\mathcal{O}|\Psi_T\rangle}{\langle R_iS_i|\Psi_T\rangle}\,,
\end{align}
where $\{R_i,S_i\}$ are spatial and spin/isospin configurations
typically sampled using the Metropolis algorithm~\cite{Metropolis:1953am},
and $\mathcal{N}$ is the (large) number of configurations in the
simulation.
In the AFDMC method, both spatial and spin/isospin degrees of freedom
are sampled during the imaginary-time propagation, the latter through
the so-called Hubbard-Stratonovich transformation.
In the GFMC approach, all possible spin/isospin configuration are
included in the trial many-body wave function, and only configurations
in coordinate space are sampled.
The above expression is valid only for observables that commute with the
Hamiltonian.
For other observables, such as radii and densities, expectation values are 
extracted from so-called mixed estimates
\begin{align}
	\label{eq:mix}
	\langle\mathcal{O}\rangle\approx2\frac{\langle\Psi_T|\mathcal{O}|\Psi(\tau)\rangle}
	{\langle\Psi_T|\Psi(\tau)\rangle}-
	\frac{\langle\Psi_T|\mathcal{O}|\Psi_T\rangle}{\langle\Psi_T|\Psi_T\rangle}\,.
\end{align}
In the above expression, the first term is the mixed estimate
(propagated wave function on one side, trial wave function on the other
side), and the second term is the variational estimate.
This relationship can be derived under the assumption that the
variational trial wave function is a good starting point, i.e. that
$|\Psi(\tau\to\infty)\rangle=|\Psi_T\rangle+|\delta\Psi_T\rangle$, with
$|\delta\Psi_T\rangle$ small. 
Then, if we calculate the expectation value of an operator between two
propagated wave functions and discard terms of $\mathcal{O}(\delta\Psi_T^2)$,
we arrive at~\cref{eq:mix}.
Additional details, including the sampling procedure and the calculation of
statistical errors, can be found, e.g. in~\cite{Ceperley:1995zz}.

The SRC scaling factors can be expressed in terms of the central
two-nucleon distribution (two-body point-nucleon
density)~\cite{Chen:2016bde}:
\begin{subequations}
	\begin{align}
		\label{eq:qmca2deut}
		a_2(A/d)&=\lim_{r\to0}\frac{2}{A}
		\frac{\rho_{2,\mathbbm{1}}(A,r)}{\rho_{2,\mathbbm{1}}(d,r)}\,,\\
		\label{eq:qmca23he}
		a_2(A/\isotope[3]{He})&=\lim_{r\to0}\frac{3}{A}
		\frac{\rho_{2,\mathbbm{1}}(A,r)}{\rho_{2,\mathbbm{1}}(\isotope[3]{He},r)}\,,
	\end{align}
\end{subequations}
where the central two-nucleon distribution is defined as
\begin{align}
	\label{eq:rho2}
	\rho_{2,\mathbbm{1}}(A,r)=\frac{1}{4\pi r^2}
	\big\langle\Psi\big|\sum_{i<j}^A\delta(r-r_{ij})\big|\Psi\big\rangle\,.
\end{align}
The normalization is such that $\rho_{2,\mathbbm{1}}(A,r)$ integrates to
the number of nucleon pairs.
\Cref{eq:rho2} involves a mixed estimate and is evaluated according
to~\cref{eq:mix}.
In this work, we ensure that the difference between the mixed and variational
estimates of the distributions is $\lesssim10\%$. 

In addition to Monte Carlo statistical errors, the use of chiral
interactions allows one to estimate the theoretical uncertainties coming
from the truncation of the chiral expansion.
In this work, we consider results for $\rho_{2,\mathbbm{1}}(A,r)$ at
leading-order (\nxlo{0}), next-to-leading-order (\nxlo{1}), and \nxlo{2},
and we estimate the truncation errors on the ratio
$X=\frac{2\rho_{2,\mathbbm{1}}(A,r)}{A\rho_{2,\mathbbm{1}}(2,r)}$ entering the 
definition of the SRC scaling factor of~\cref{eq:qmca2deut,eq:qmca23he}
following~\cite{Epelbaum:2014efa}:
\begin{align}
	\label{eq:err}
	\Delta X^\text{\nxlo{2}}=&\max(Q^4|X^{\text{\nxlo{0}}}|,
	Q^2|X^\text{\nxlo{1}}-X^\text{\nxlo{0}}|,
	Q|X^\text{\nxlo{2}}-X^\text{\nxlo{1}}|)\,,
\end{align}
where we take $Q=m_\pi/\Lambda_b$ with $m_\pi\approx140\,\rm MeV$ and
$\Lambda_b=600\,\rm MeV$, as in~\cite{Lonardoni:2017hgs,Lonardoni:2018nob}.

Auxiliary field diffusion Monte Carlo calculations for nuclei employing
local chiral interactions have been carried out up to
$A=16$~\cite{Lonardoni:2017hgs,Lonardoni:2018nob,Lonardoni:2018sqo}.
Preliminary results for heavier systems suggest that improved wave
functions are necessary to obtain ground-state properties with the same
accuracy as for lighter systems.
However, such a prescription will increase the computational cost by a
factor proportional to $A^2$ (see Ref.~\cite{Lonardoni:2018nob} for
details), making calculations for $A\gtrsim 20$ no longer feasible.
One way to move beyond oxygen is to use a simplified interaction,
capable of capturing most of the ground-state physics of nuclei, for
which the employed wave function still gives an accurate description of
larger nuclei, thus maintaining the good computational scaling of the
current implementation of the AFDMC algorithm. 
We consider the phenomenological two-body Argonne $v_4'$ potential
($\text{AV4}'$)~\cite{Wiringa:2002ja}, a simplified version of the more
sophisticated Argonne $v_{18}$ (AV18) potential~\cite{Wiringa:1994wb},
obtained by reprojecting the full potential onto the first four operator
channels in order to preserve the phase shifts of lower partial waves and
the deuteron binding energy.
We note that this potential is very simple and excludes, for example,
tensor forces.
The Coulomb interaction is, however, still included.
Such a potential typically overbinds light nuclei~\cite{Wiringa:2002ja}.
The inclusion of a repulsive three-body force can be used to compensate
for the excessive attraction.
As done in other
works~\cite{Lonardoni:2013rm,Lonardoni:2013gta,Lonardoni:2017uuu}, we
consider the central component of the Urbana IX (UIX)
interaction~\cite{Pudliner:1997ck} as a source of repulsion.
In the following, this simplified potential will be referred to as 
$\text{AV4}'+\text{UIX}_\text{c}$.

\section{Results}
\label{sec:res}

\subsection{Fitting $a_2$}
In~\cite{Chen:2016bde}, the SRC scaling factors were obtained by
taking the limit $r\to0$ of the ratio of two-body distributions as
in~\cref{eq:qmca2deut,eq:qmca23he}.
However, this is precisely the region where the Monte Carlo statistical
uncertainties become large, see, e.g. \cref{fig:a2_6he_12c_40ca}.
Nuclear potentials are generally repulsive at short distances, and
therefore the likelihood of finding two nucleons at small
separations is small, giving rise to large statistical uncertainties
as $r\to0$.)
In this work, we exploit the fact that, as pointed out in~\cite{Chen:2016bde}, 
in EFT, ``locality'' means a shorter distance
than the resolution scale.
Thus, we expect that we can replace $r\to0$
in~\cref{eq:qmca2deut,eq:qmca23he} by simply smearing in some region
$r<R$, where $R$ is set by the cutoff scale $R_0$ (but $R$ is not
necessarily equal to $R_0$), and still obtain the same $a_2$ value.
Therefore, we fit a horizontal line to the ratio of two-body distributions
$2\rho_{2,\mathbbm{1}}(A,r)/A\rho_{2,\mathbbm{1}}(d,r)$ and
$3\rho_{2,\mathbbm{1}}(A,r)/A\rho_{2,\mathbbm{1}}(\isotope[3]{He},r)$ in
the region with $0\le r\le R$, and we take $R=0.7\,\text{fm}$.
This region is chosen as the empirical region where the expected plateau
sets in.
(Note that for the systems \isotope[3]{H}, \isotope[3]{He}, and
\isotope[4]{He}, the results from this linear fit agree with our
previous results from~\cite{Chen:2016bde} using the limit $r\to0$).
We have further checked that varying $R$ from $0.4$ to $1.0\,\text{fm}$ 
makes a 1\%--3\% difference in our extracted values of
$a_2$ for the local chiral interactions.
For results with the simplified $\text{AV4}'+\text{UIX}_\text{c}$
and the $\text{AV18}+\text{UIX}$ potentials, varying $R$ in this range
makes a $\sim10\%$ difference (up to 13\% for \isotope[40]{Ca}).
For the phenomenological potentials, we use this variation as an
estimate for the systematic uncertainty coming from the fit of $a_2$.

If each of the discrete values $\{\rho_{2,\mathbbm{1}}(A,r_i)\}$ obtained
from Monte Carlo calculations were equally likely, this procedure would
be entirely equivalent to taking the average in the region $0\le r\le R$.
However, as discussed above, the statistical uncertainties in
$\rho_{2,\mathbbm{1}}(A,r)$ grow rapidly as $r\to0$.
In short, our fitting problem is heteroskedastic, and therefore, we use a
weighted linear least squares fitting procedure, 
\begin{align}
	a_2(A)=(X^TWX)^{-1}X^TWy\,,
\end{align}
where we take the weight
matrix diagonal and equal to the inverse of the Monte Carlo variances for
each point:
\begin{align}
	W_{ii}\to\frac{1}{\sigma_i^2}\,.
\end{align}
In our case $X_{ij}$ reduces to a vector of $1$'s, and the $\{y_i\}$ are
the set of values $\{a_2(r_i)\}$ from the Monte Carlo simulations.
Then our procedure amounts to
\begin{align}
	a_2(A)=\frac{\sum_{i=1}^\mathcal{M}\frac{1}{\sigma_i^2}
	\frac{2\rho_{2,\mathbbm{1}}(A,r_i)}{A\rho_{2,\mathbbm{1}}(2,r_i)}}
	{\sum_{i=1}^\mathcal{M}\frac{1}{\sigma_i^2}}\,,
\end{align}
where $\mathcal{M}$ is taken such that $r_i\in[0.0,0.7]\,\rm fm$.

\subsection{Results for selected nuclei up to \isotope[40]{Ca}}
\begin{figure}[t]
	\centering
	\includegraphics[width=0.8\columnwidth]{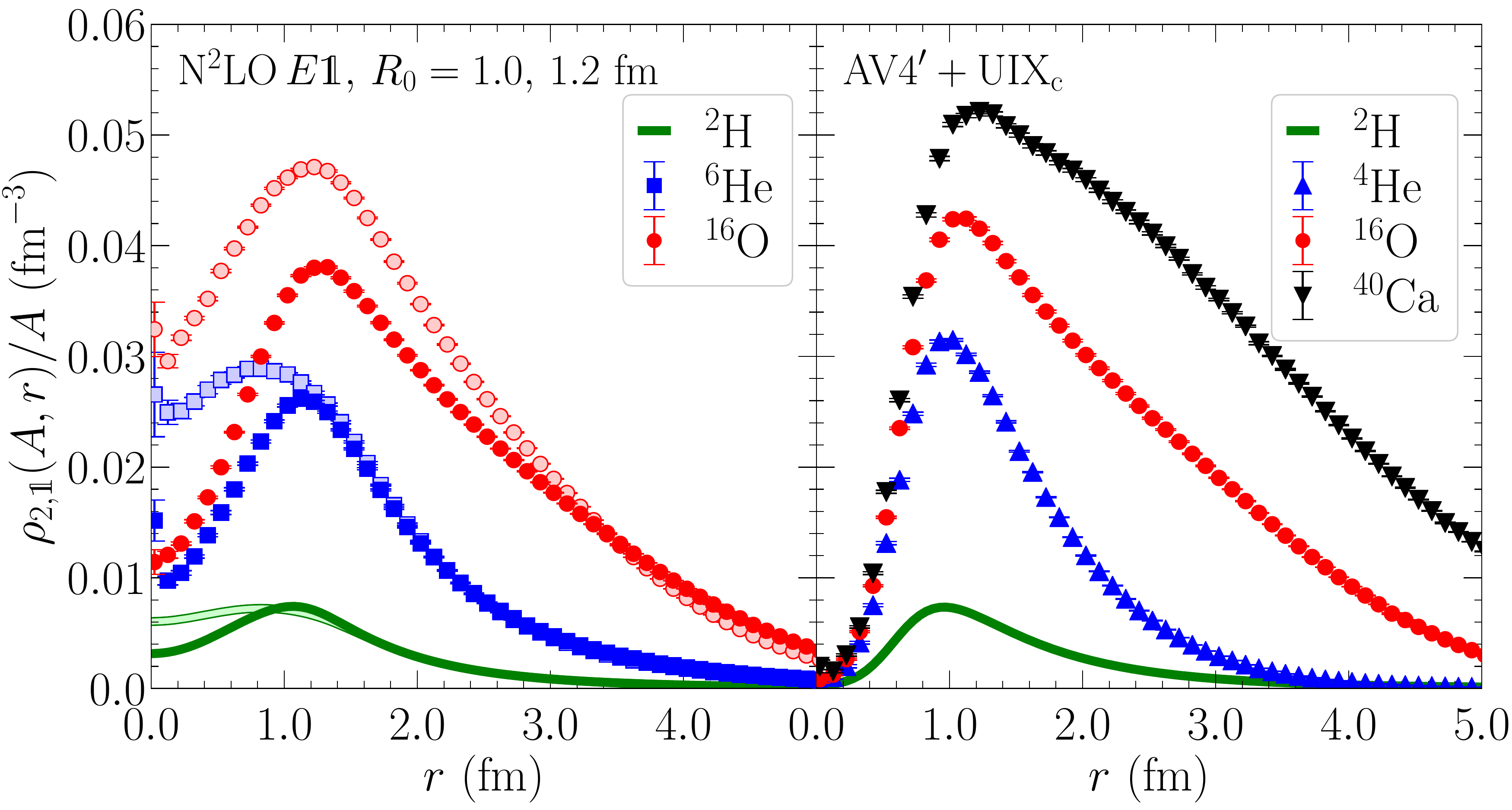}
	\caption[]{\label{fig:gofr}
	Scaled two-nucleon distributions at \nxlo{2} for \isotope[2]{H},
	\isotope[6]{He}, and \isotope[16]{O} for the $3N$ parameterization
	$E\mathbbm{1}$ (left panel).
	The darker (lighter) colors correspond to $R_0=1.0\ (1.2)\,\rm fm$.
	The right panel shows the scaled two-nucleon distributions for the
	$\text{AV4}'+\text{UIX}_\text{c}$ potential for \isotope[2]{H}, 
	\isotope[4]{He}, \isotope[16]{O}, and \isotope[40]{Ca}.
	}
\end{figure}
We first present results for the two-body distributions~\cref{eq:rho2}
for selected nuclei up to $\isotope[40]{Ca}$ in~\cref{fig:gofr}.
The left panel shows results for the deuteron, \isotope[6]{He}, and
\isotope[16]{O} using local chiral interactions at \nxlo{2} with the
$E\mathbbm{1}$ parameterization of the $3N$ interaction and both cutoffs
$R_0=1.0,\,1.2\,\rm fm$, whereas the right panel shows results for the
deuteron, \isotope[4]{He}, \isotope[16]{O}, and \isotope[40]{Ca} for the
simplified nuclear potential $\text{AV4}'+\text{UIX}_\text{c}$.
The figure shows the definite scheme and scale dependence of these
distributions.
This is especially clear in the left panel where the distributions are
calculated in an EFT framework at two cutoff scales.
The softer cutoff $R_0=1.2\,\rm fm$ resembles more a mean-field calculation
at short distances where what are typically referred to as SRCs are
reduced, leading to a higher probability to find two nucleons separated
by very short distances $r\lesssim1.0\,\rm fm$.
The right panel, which utilizes phenomenological potentials where the
effective cutoff is much harder (though a particular value is not
identified) shows a significantly lower probability to find a pair of
nucleons separated by $r\lesssim1.0\,\rm fm$.

\begin{figure}[t]
	\centering
	\includegraphics[width=\columnwidth]{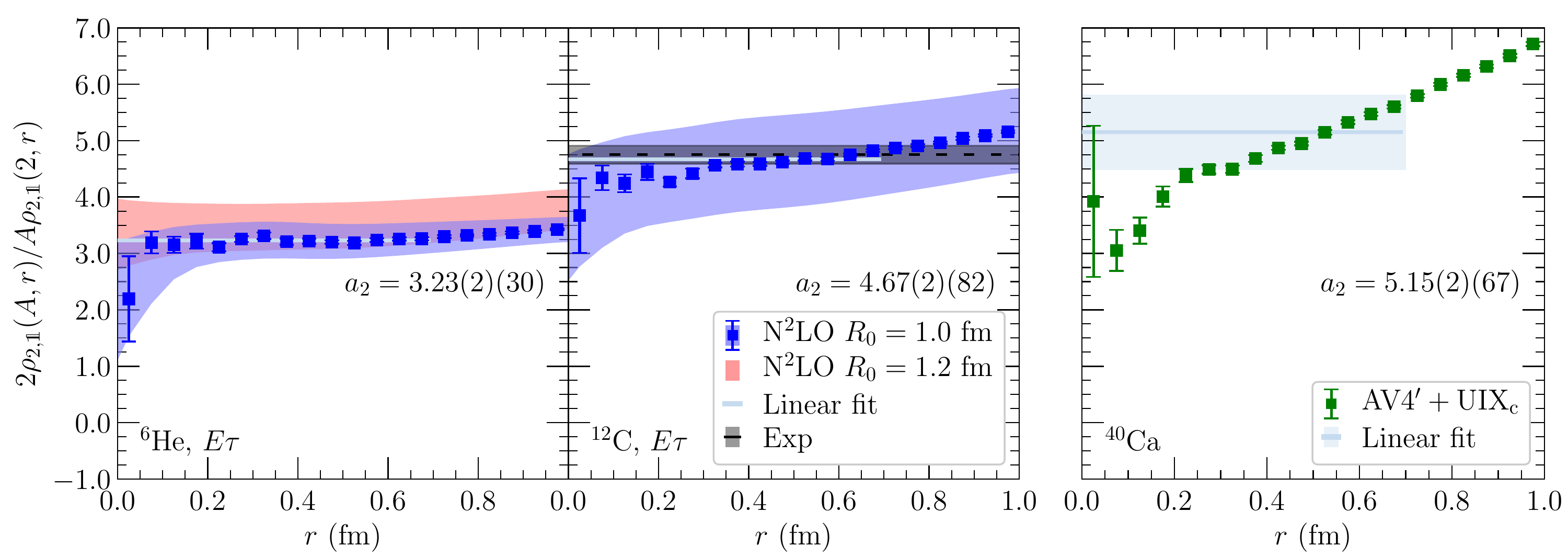}
	\caption[]{\label{fig:a2_6he_12c_40ca}
	Three examples of the extraction of the SRC scaling factor $a_2(A/d)$
	from Monte Carlo results.
	The left two panels show results for the local chiral interactions at
	\nxlo{2} with the $E\tau$ parameterization of the $3N$ force for
	\isotope[6]{He} and \isotope[12]{C}.
	The right panel shows results for the
	$\text{AV4}'+\text{UIX}_\text{c}$ potential for \isotope[40]{Ca}.
	For the chiral interactions, we indicate the combined statistical and
	chiral truncation uncertainty estimates as the blue and red bands.
	For the phenomenological potentials (right panel) we indicate the
	uncertainty in the fit by the light blue band.
	In each case, as described in more detail in the text, we fit a
	horizontal line to the AFDMC results weighted by the Monte Carlo
	statistical uncertainties in the region $0\le r\le0.7\,\rm fm$.
	The values extracted for $a_2$ using this procedure are shown in each
	panel including uncertainties.
	For \isotope[12]{C} the experimental value~\protect\cite{Hen:2012fm} with
	uncertainties is shown as the black dashed line with the gray band.
	}
\end{figure}

In contrast to the two-body distributions shown
in~\cref{fig:gofr}, their ratios to the deuteron and
\isotope[3]{He} two-body distributions,
i.e.~\cref{eq:qmca2deut,eq:qmca23he} are largely scheme and scale
independent.
In~\cref{fig:a2_6he_12c_40ca}, we show the ratio $a_2(A/d)$ at short
internucleon distances $0\le r\le1.0$~fm for \isotope[6]{He} and
\isotope[12]{C} using chiral EFT interactions at \nxlo{2} with the
$E\tau$ parameterization of the $3N$ interaction (left and middle
panels) and for \isotope[40]{Ca} using the simplified
$\text{AV4}'+\text{UIX}_\text{c}$ potential (right panel).
For \isotope[6]{He} and \isotope[12]{C}, we show the Monte Carlo results
with statistical uncertainties using $R_0=1.0\,\rm fm$ as the blue squares
with error bars, while we use green squares for \isotope[40]{Ca} using the simplified
$\text{AV4}'+\text{UIX}_\text{c}$ potentials.
The blue (red) band represents the combined statistical and systematic
uncertainties coming from the truncation of the chiral expansion for the
$R_0=1.0\,\rm fm$ ($1.2\,\rm fm$) cutoff.
The light blue band in the right panel shows the uncertainty in the fit
for \isotope[40]{Ca}.
The light blue horizontal lines indicate the weighted linear fits to the
Monte Carlo results as described above.
Of the three cases shown here, there is currently only an experimental
result with which to compare for \isotope[12]{C}. This is shown in the middle
panel as the black dashed line with the gray band representing the
experimental uncertainty.

\begin{figure}[t]
	\centering
	\includegraphics[width=0.495\columnwidth]{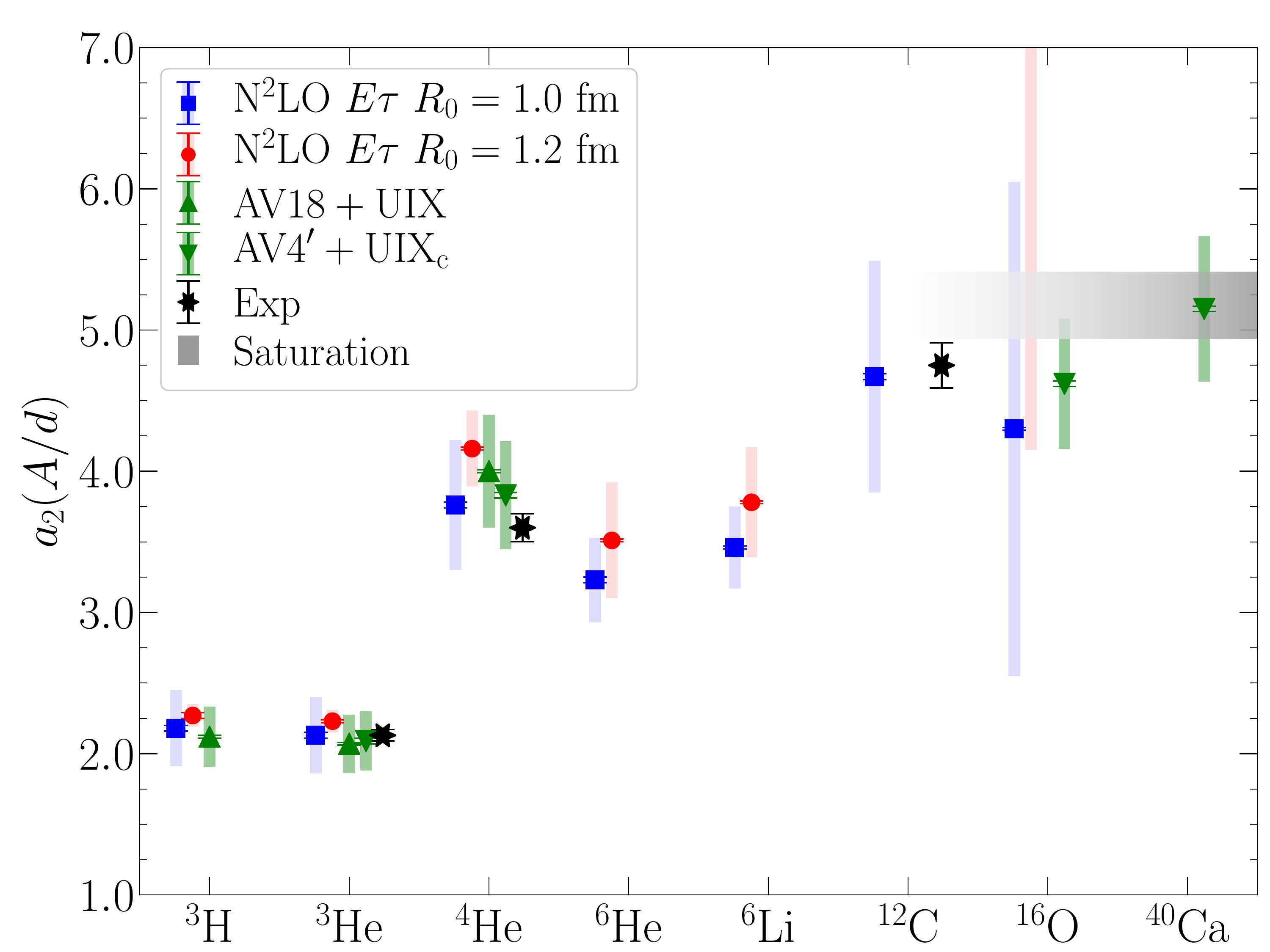}
	\includegraphics[width=0.495\columnwidth]{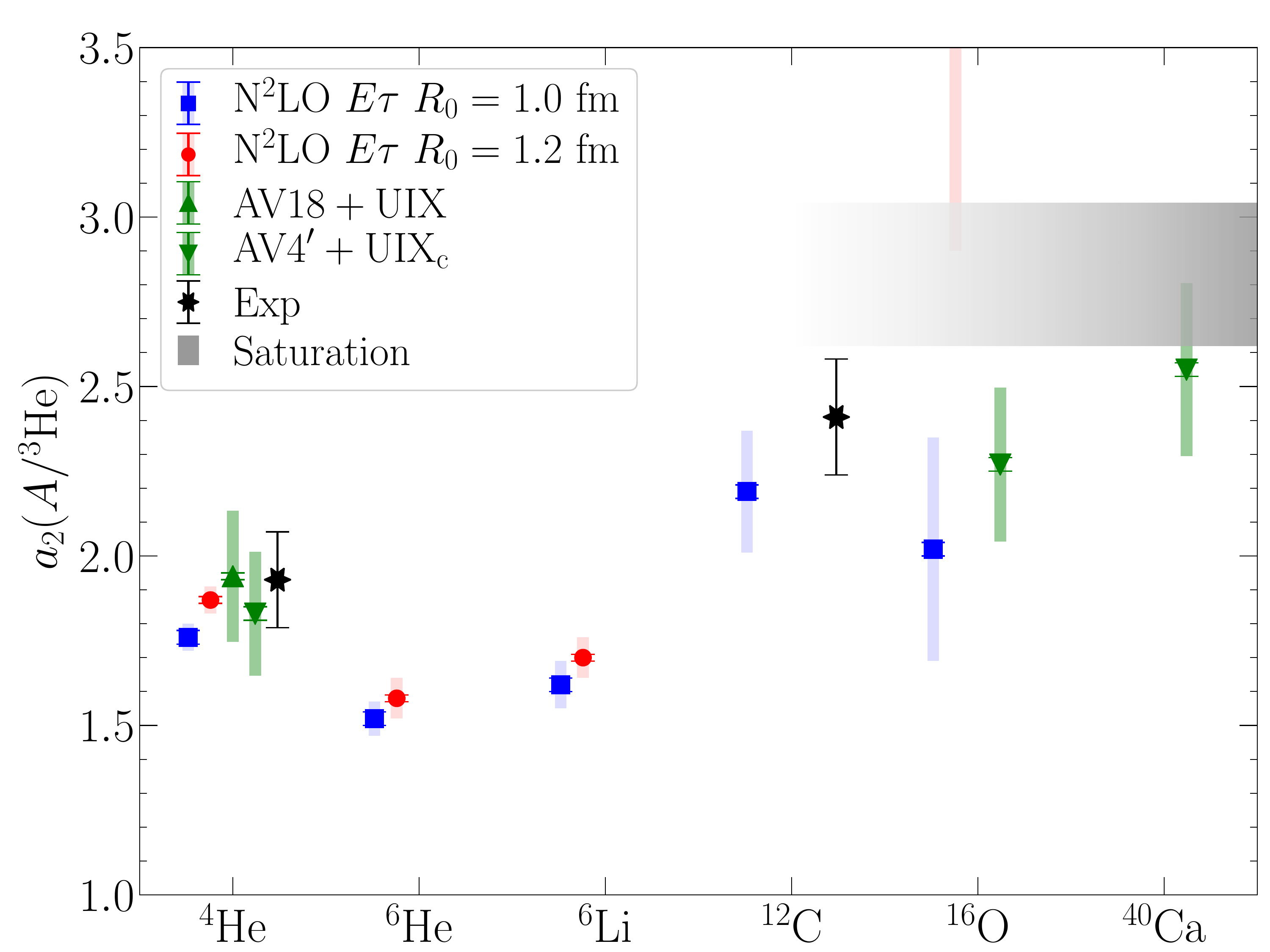}
	\caption[]{\label{fig:a2}
	Short-range-correlation scaling factors $a_2$ for selected nuclei
	from $A=3$ up to $A=40$ calculated with respect to the deuteron (left
	panel) and \isotope[3]{He} (right panel).
	Results for the chiral interactions at \nxlo{2} (with the $E\tau$
	parameterization of the $3N$ force) for cutoff $R_0=1.0\ (1.2)\,\rm fm$ 
	are shown as the blue squares (red circles).
	We also show results for the $\text{AV18}+\text{UIX}$ potentials
	(green upward-pointing triangles) as well as the simplified 
	$\text{AV4}'+\text{UIX}_\text{c}$ potentials (green downward-pointing triangles).
	The black stars in the left (right) panel are the experimental values from
	Ref.~\protect\cite{Hen:2012fm} (Ref.~\protect\cite{Egiyan:2005hs}).
	The gray bands represent the expected range of values at which $a_2$ saturates,
	based on measurements for \isotope[63]{Cu} and \isotope[197]{Au}~\protect\cite{Hen:2012fm}
	(also reported in~\cref{tab:a2}) in the left panel and based on
	measurements for \isotope[56]{Fe}~\protect\cite{Egiyan:2005hs} (also reported
	in~\cref{tab:a2}) in the right panel.
	The dark error bars (typically smaller than the symbols) represent the
	Monte Carlo statistical uncertainties. 
	The lighter bands show the overall systematic uncertainties, both
	associated with the truncation of the chiral expansion at \nxlo{2} 
	as computed using~\cref{eq:err} for local chiral interactions, and 
	coming from the fit of $a_2$ for the phenomenological potentials
	(see the text for more details).
	}
\end{figure}

While~\cref{fig:a2_6he_12c_40ca} illustrates the method by which we
extract the SRC scaling factors, in \cref{fig:a2} we show our
main predictions.
The left panel shows results for $a_2(A/d)$ for selected nuclei from
\isotope[3]{H} up to \isotope[40]{Ca}.
The blue squares (red circles) show the results for chiral interactions
up to \nxlo{2} with the $E\tau$ parameterization of the $3N$ interaction
and the cutoff $R_0=1.0\,\rm fm$ ($R_0=1.2\,\rm fm$).
The green upward-pointing (downward-pointing) triangles show the results
using the $\text{AV18}+\text{UIX}$ ($\text{AV4}'+\text{UIX}_\text{c}$)
potentials.
The black stars show the experimental results from~\cite{Hen:2012fm},
where available.
The light colored bands show the systematic uncertainties 
stemming from the truncation of the chiral expansion at \nxlo{2} for the $R_0=1.0\,\rm fm$
($R_0=1.2\,\rm fm$) cutoff (where available), and coming from the fit of
$a_2$ in the case of the phenomenological potentials.
The gray region appearing at large $A$ represents an expected saturation
region taken as the difference, including uncertainties, between
$a_2(\isotope[197]{Au}/d)=5.16(22)$ and
$a_2(\isotope[63]{Cu}/d)=5.21(20)$~\cite{Hen:2012fm}, i.e. we estimate
$\lim_{A\to\infty}a_2(A/d)\sim4.94\text{--}5.41$.
Similarly, the right panel of \cref{fig:a2} shows results for
$a_2(A/\isotope[3]{He})$ for selected nuclei from \isotope[4]{He} up to
\isotope[40]{Ca} using the same color and symbol scheme as in the left
panel.
Note that the gray saturation region is provided by a single
experimental value at large $A$: namely,
$a_2(\isotope[56]{Fe}/\isotope[3]{He})=2.83(3)(18)$~\cite{Egiyan:2005hs}.
(\cref{tab:a2} collect
these and more results using both parametrizations $E\tau$ and
$E\mathbbm{1}$ of the $3N$ interaction and both cutoffs
$R_0=1.0$,~1.2~fm, as well as results for phenomenological potentials and
experimental results).

The results in~\cref{fig:a2} compare very well with experimental
values, where available.
In particular, we find 0.0\%, 4.4\%, and 1.7\% relative agreement
between our results for $a_2(A/d)$ using chiral interactions at \nxlo{2}
with the cutoff $R_0=1.0\,\rm fm$ and experiment for \isotope[3]{He},
\isotope[4]{He}, and \isotope[12]{C}, respectively.
Results using the softer cutoff $R_0=1.2\,\rm fm$ are typically higher than
for the lower cutoff by $\sim5\text{--}10\%$ (an exception occurs for
\isotope[16]{O}, where the softer interaction with the $E\tau$
parametrization has already been found to exhibit significant
overbinding~\cite{Lonardoni:2018nob}), but are always within the
estimated systematic uncertainties.
It is also interesting to note that the predicted values for $a_2$ for
the $A=6$ systems fall below the values for \isotope[4]{He}, placing them
between \isotope[3]{He} and \isotope[4]{He} along the fitted line
in~\cref{fig:drdx_a2}.
As suggested in~\cite{Seely:2009gt}, what appears to
dictate the strength of the EMC effect (and therefore the height of the
SRC scaling plateaus through the EMC-SRC linear relation) is the
local nuclear density.
Given that \isotope[4]{He} is such a compact nucleus, and that both
\isotope[6]{Li} and \isotope[6]{He} can be thought of as $\alpha$
particles with additional nucleons ``orbiting,'' one might expect that
the strong attraction of the $\alpha$ core to the orbiting nucleons
would tend to lower the local central two-nucleon density.
These predictions for \isotope[6]{Li} could be tested already using
existing experimental setups for $(e,e')$ inclusive scattering in QE
kinematics at Jefferson Lab.
For \isotope[6]{He}, these predictions could be tested at future rare
isotope facilities such as the Facility for Antiproton and Ion Research
with experiments in inverse kinematics using a
\isotope[6]{He} beam on a proton target inducing $(p,2p)$ reactions.

\begin{sidewaystable}
\vspace{15cm}
	\centering
	\caption[]{
	Results for the SRC scaling factor $a_2(A/d)$ (upper table) and $a_2(A/\isotope[3]{He})$ 
	(lower table) obtained via a linear fit
	to the Monte Carlo results for different nuclear interactions (see text
	for details).
	Both statistical (first) and systematic (second) uncertainties are
	reported in the parentheses.
	The latter include both the uncertainty coming from the fit of $a_2$ and
	the uncertainty	associated with the truncation of the chiral expansion 
	(for local chiral interactions).
	$\text{AV18}+\text{UIX}$ results are from GFMC
	calculations~\protect\cite{Chen:2016bde}, while the other results are obtained
	using the AFDMC method.
	The last column shows the available experimental data from~\protect\cite{Hen:2012fm,Egiyan:2005hs}.\\
	}
	\begin{tabular}{lccccccc}
	System & \multicolumn{2}{c}{$\text{\nxlo{2}}\ E\tau$} &
	\multicolumn{2}{c}{$\text{\nxlo{2}}\ E\mathbbm{1}$} &
	\multicolumn{1}{l}{$\text{AV18}+\text{UIX}$} &
	\multicolumn{1}{l}{$\text{AV4}'+\text{UIX}_\text{c}$} & Exp \\ 
	& \multicolumn{1}{l}{$R_0=1.0\text{ fm}$} & \multicolumn{1}{l}{$R_0=1.2\text{ fm}$} & \multicolumn{1}{l}{$R_0=1.0\text{ fm}$} & \multicolumn{1}{l}{$R_0=1.2\text{ fm}$} & & & \\
	\hline                                                                                                                                   
	\isotope[3]{H}    & $2.18(2)(27)  $ & $2.27(2)(8)   $ & $2.15(2)(28)  $ & $2.46(2)(8)   $ & $2.12(1)(22)$ & $-          $ & $-          $ \\
	\isotope[3]{He}   & $2.13(2)(27)  $ & $2.23(1)(8)   $ & $2.10(2)(28)  $ & $2.38(2)(8)   $ & $2.07(1)(21)$ & $2.09(2)(21)$ & $2.13(4)    $ \\ 
	\isotope[4]{He}   & $3.76(2)(46)  $ & $4.16(1)(27)  $ & $3.77(2)(46)  $ & $5.31(2)(27)  $ & $4.00(1)(40)$ & $3.83(2)(39)$ & $3.60(10)   $ \\
	\isotope[6]{He}   & $3.23(2)(30)  $ & $3.51(1)(41)  $ & $3.14(1)(30)  $ & $4.04(2)(41)  $ & $-          $ & $-          $ & $-          $ \\
	\isotope[6]{Li}   & $3.46(1)(29)  $ & $3.78(1)(39)  $ & $3.33(1)(29)  $ & $4.18(2)(39)  $ & $-          $ & $-          $ & $-          $ \\
	\isotope[12]{C}   & $4.67(2)(82)  $ & $-            $ & $-            $ & $-            $ & $-          $ & $-          $ & $4.75(16)   $ \\
	\isotope[16]{O}   & $4.30(1)(1.75)$ & $8.55(1)(4.40)$ & $4.02(1)(1.75)$ & $5.47(1)(4.40)$ & $-          $ & $4.62(2)(47)$ & $-          $ \\
	\isotope[40]{Ca}  & $-            $ & $-            $ & $-            $ & $-            $ & $-          $ & $5.15(2)(67)$ & $-          $ \\
	\isotope[63]{Cu}  & $-            $ & $-            $ & $-            $ & $-            $ & $-          $ & $-          $ & $5.21(20)   $ \\
	\isotope[197]{Au} & $-            $ & $-            $ & $-            $ & $-            $ & $-          $ & $-          $ & $5.16(22)   $ \\
	\hline
	\isotope[4]{He}   & $1.76(2)(4) $   & $1.87(1)(4)   $ & $1.80(2)(4)   $ & $2.23(2)(9)   $ & $1.83(2)(19)$ & $1.94(1)(20)$ & $1.93(2)(14)$ \\	
	\isotope[6]{He}   & $1.52(2)(5) $   & $1.58(1)(6)   $ & $1.50(2)(4)   $ & $1.70(2)(7)   $ & $-          $ & $-          $ & $-          $ \\	
	\isotope[6]{Li}   & $1.62(2)(7) $   & $1.70(1)(6)   $ & $1.59(2)(6)   $ & $1.76(2)(6)   $ & $-          $ & $-          $ & $-          $ \\	
	\isotope[12]{C}   & $2.19(2)(18)$   & $-            $ & $-            $ & $-            $ & $-          $ & $-          $ & $2.41(2)(17)$ \\	
	\isotope[16]{O}   & $2.02(2)(33)$   & $3.91(2)(1.01)$ & $1.91(2)(33)  $ & $2.27(2)(1.01)$ & $-          $ & $2.27(2)(23)$ & $-          $ \\	
	\isotope[40]{Ca}  & $-          $   & $-            $ & $-            $ & $-            $ & $-          $ & $2.55(2)(33)$ & $-          $ \\	
	\isotope[56]{Fe}  & $-          $   & $-            $ & $-            $ & $-            $ & $-          $ & $-          $ & $2.83(3)(18)$ \\	
	\end{tabular}
	\label{tab:a2}
\end{sidewaystable}

\begin{table}[tb]
	\centering
	\caption[]{\label{tab:av4c+uce}
	Binding energies (in MeV) and charge radii (in fm) for $A=4,16,40$
	with the $\text{AV4}'+\text{UIX}_\text{c}$ potential.
	Energy results are from the AFDMC unconstrained
	evolution~\protect\cite{Lonardoni:2018nob}.
	Experimental results are shown for comparison.\\
	}
	\begin{tabular}{lcccr}
	\isotope[A]{Z} & \multicolumn{1}{c}{$E_\text{AFDMC}$} &
	\multicolumn{1}{c}{$E_\text{Exp}$} &
	\multicolumn{1}{c}{$r_\text{ch}^\text{AFDMC}$} &
	\multicolumn{1}{c}{$r_\text{ch}^\text{Exp}$} \\ 
	\hline
	\isotope[4]{He}  & $-26.00(2)$ & $-28.296$  & $1.74(1)$ & $1.680(4)$~\protect\cite{Sick:2008} \\
	\isotope[16]{O}  & $-113(2)$   & $-127.619$ & $2.61(6)$ & $2.699(5)$~\protect\cite{Angeli:2013epw} \\
	\isotope[40]{Ca} & $-321(3)$   & $-342.052$ & $3.25(8)$ & $3.478(2)$~\protect\cite{Angeli:2013epw}\\
	\end{tabular}
\end{table}

We also make predictions for \isotope[16]{O} and \isotope[40]{Ca}
in~\cref{fig:a2,tab:a2}.
While the latter is only calculated using the simplified
phenomenological potential $\text{AV4}'+\text{UIX}_\text{c}$, our
expectation based on calculations for light systems with $3\le A\le16$
is that this rather simplified Hamiltonian is capturing most of the
important SRC physics:
this can be seen by comparing the results using the realistic chiral
EFT interactions at \nxlo{2} with $R_0=1.0\,\rm fm$ (blue squares
in~\cref{fig:a2}) with the results using
$\text{AV4}'+\text{UIX}_\text{c}$ (green downward-pointing triangles).
The relative agreement between the results is 1.9\%, 1.9\%, and 7.4\%
for \isotope[3]{He}, \isotope[4]{He}, and \isotope[16]{O}, respectively.
We also refer the reader to~\cref{tab:av4c+uce}: Both the binding
energies and radii for \isotope[4]{He}, \isotope[16]{O}, and
\isotope[40]{Ca} are reasonably well reproduced using
$\text{AV4}'+\text{UIX}_\text{c}$.
Nevertheless, given the relative agreement between our chiral
interactions at \nxlo{2} with the cutoff $R_0=1.0\,\rm fm$ and the simplified
potential $\text{AV4}'+\text{UIX}_\text{c}$, and the slight systematic
underbinding of the latter, we assign a conservative uncertainty to our
$\text{AV4}'+\text{UIX}_\text{c}$ calculations, e.g.
$a_2(\isotope[40]{Ca}/d)=5.15(67)$ and
$a_2(\isotope[40]{Ca}/\isotope[3]{He})=2.55(33)$.
This 13\% can be justified from our study of the sensitivity of the
extracted $a_2$ to the chosen region $0\le r\le R$.

\section{Summary}
\label{sec:sum}
In this work, we have used DMC algorithms, namely the
GFMC and AFDMC methods, to calculate the SRC scaling factors $a_2(A/d)$
and $a_2(A/\isotope[3]{He})$ for nuclei from $A=3$ to $A=40$.
We have reviewed in detail the derivation of $a_2$ from EFT, arguing that
isovector corrections are very small.
We have then shown that fitting a constant to the ratio of two-body
central densities in some empirical region $0\le r\le R$ reproduces the
values from our previous work~\cite{Chen:2016bde} and provides a
reliable method to extract SRC scaling factors.
Where experimental values exist, our calculations agree very well
using both chiral EFT interactions at \nxlo{2} and phenomenological
potentials, including the simplified $\text{AV4}'+\text{UIX}_\text{c}$
potential, providing further evidence of the value of the novel framework
first proposed in~\cite{Chen:2016bde}.
We also show the first \textit{ab initio} predictions for SRC scaling
factors for \isotope[6]{Li}, \isotope[6]{He}, \isotope[16]{O}, and
\isotope[40]{Ca}.
These predictions could be tested in future experiments, offering
intriguing insights into the evolution of SRC scaling factors with the
nuclear mass $A$.
Our framework may also shed light on the proposed, but so far elusive,
$3N$ SRC scaling.
This topic is currently being investigated and we leave it for future
work.

\ack{
We thank N~Fomin for providing us with the data from~\cite{Fomin:2011ng} 
and T~Aumann, H-W~Hammer, K~Hebeler,
O~Hen, and I~Tews for valuable discussions. 
The work of JL and AS was supported by the ERC Grant No.~307986
STRONGINT and the BMBF under Contract No.~05P18RDFN1.
The work of DL was supported by the US Department of Energy, Office
of Science, Office of Nuclear Physics, under Contract No.~DE-SC0013617,
and by the NUCLEI SciDAC program.
The work of JC and SG was supported by the NUCLEI SciDAC program, by
the US Department of Energy, Office of Science, Office of Nuclear
Physics, under contract No.~DE-AC52-06NA25396, and by the LDRD program
at LANL.
SG was also supported by the DOE Early Career research Program.
J-W C is partly supported by the Ministry of Science and Technology,
Taiwan, under Grant No.~105-2112-M-002-017-MY3 and the Kenda Foundation.
WD is supported by the US Department of Energy under grant DE-SC0011090.
WDUS is also supported within the framework of the TMD Topical
Collaboration of the US Department of Energy, Office of Science,
Office of Nuclear Physics, and by the SciDAC4 award DE-SC0018121.
Computational resources have been provided by the Lichtenberg high
performance computer of the TU Darmstadt, 
by the Los Alamos National Laboratory Institutional Computing Program, 
which is supported by the US Department of Energy 
National Nuclear Security Administration under Contract No.~89233218CNA000001
and by the
National Energy Research Scientific Computing Center (NERSC), which is
supported by the US Department of Energy, Office of Science, under
contract No.~DE-AC02-05CH11231.
}

\providecommand{\newblock}{}

\end{document}